**Two Roads Diverged: A Semantic Network Analysis of *Guanxi* on Twitter**

Pu Yan and Taha Yasseri

Taha.yasseri@oii.ox.ac.uk
Oxford Internet Institute, University of Oxford, 1 St Giles OX13JS Oxford, UK

**Abstract**

*Guanxi,* roughly translated as "social connection", is a term commonly used in the Chinese language. In this research, we employed a linguistic approach to explore popular discourses on *Guanxi*. Although sharing the same Confucian roots, Chinese communities inside and outside Mainland China have undergone different historical trajectories. Hence, we took a comparative approach to examine *guanxi* in Mainland China and in Taiwan, Hong Kong, and Macau (TW-HK-M). Comparing *guanxi* discourses in two Chinese societies aims at revealing the divergence of *guanxi* culture. The data for this research were collected on Twitter over a three-week period by searching tweets containing *guanxi* written in Simplified Chinese characters (关系) and in Traditional Chinese characters (關係). After building, visualising, and conducting community detection on both semantic networks, two *guanxi* discourses were then compared in terms of their major concept sub-communities. This research aims at addressing two questions: Has the meaning of *guanxi* transformed in contemporary Chinese societies? And how do different socio-economic configurations affect the practice of *guanxi*? Results suggest that *guanxi* in interpersonal relationships has adapted to a new family structure in both Chinese societies. In addition, the practice of *guanxi* in business varies in Mainland China and in TW-HK-M. Furthermore, an extended domain was identified where *guanxi* is used in a macro-level discussion of state relations. Network representations of the *guanxi* discourses enabled reification of the concept and shed lights on the understanding of social connections and social orders in contemporary China.

# 1. Introduction

*Guanxi* is one of the key concepts scholars often use to describe Chinese societal features. It is the Romanization (*pinyin*) spelling of the Chinese character "关系" (in simplified Chinese) or "關係" (in traditional Chinese). It can be translated as "social connections" or "personal relationships". However, in Chinese language, the word entails much richer meanings than social connections, and is bound to notions such as obligation and loyalty to ascribed social groups. As specified by the Oxford Dictionary of English (2010), *guanxi* is "the relationship that facilitates business and other dealings". This definition emphasises the functional role of *guanxi* in Chinese interpersonal relationships.

*Guanxi* relations sometimes refer to an exclusive and yet politically influential social clique that shares similar *guanxi* ties. One recent example of the *guanxi* network in politics comes from China's continuing struggle against corruption and conspiracy. A group of government officials were indicted for corruption earlier this year. According to reports on Chinese state media, this network was established among *tongxiang* (同乡), people from the same hometown, forming a "protective umbrella" to cover up bribery and embezzlement, leading to a "systematic corruption" (Yi, 2014).

Today, Confucian culture still exists in Chinese societies around the world, including Mainland China, Taiwan, Hong Kong, Macau, and overseas Chinese communities (Jensen, 1997; Barmé, 2000). *Guanxi* culture, the social norms extended from traditional Confucian teaching, still exists in contemporary Chinese societies in the Mainland and in the greater Chinese regions, including Taiwan, Hong Kong, and Macau. However, these societies have gradually separated into distinct societies since the 1950s (Lin and Ho, 2009).

In Mainland China, Chinese traditional ethics and moral principles were first criticized and revolutionized with the establishment of People's Republic of China (PRC) in 1949 and the Cultural Revolution from 1960s to 1970s. The ideological revolution was reflected in the Party's avocation of simplified Chinese (*jiantizi* 简体字) in formal printing and writing. However, the post-Mao era has seen the decline of Marxism and the revival of traditional values. culture (Bell, 2010). The political rhetoric and the social orders that were valued in Confucianism, such as loyalty, family obligations, harmony, and concerns for others, have been once again emphasised in official and mainstream ideology.

Meanwhile, the greater Chinese communities outside Mainland China (for example, Taiwan, Hong Kong and Macau) demonstrate distinct socio-economic and cultural features due to having had different historical trajectories. The democratization of Taiwan took place in the 1980s and 1990s, which encouraged the public to participate in the political system and diversified the ideological system in Taiwan. As former colonies of Britain and Portugal, Hong Kong and Macau had been exposed to foreign cultures and ethics throughout their colonial history. Yet, Chinese culture and traditions still have an impact on the ethical and moral rules in Hong Kong and Macau (Wong, 1986). Taiwan, Hong Kong, and Macau did not adopt simplified Chinese; thus, traditional Chinese still dominates both the informal and formal writing systems. As Chinese societies adopt modern values, questions have been raised within academia on whether or not Chinese traditional culture still has profound influence on Chinese contemporary interpersonal relationships, to what extend it has adapted to social changes that come with Chinese modernization, and how it has diverged in Chinese communities that have demonstrated different socio-economic backgrounds.

## 1.1 Related Work

### 1.1.1 Conceptualising *Guanxi*

Early scholars studying *guanxi* culture traced its roots back to Chinese social structure. Chinese sociologist, Fei Xiaotong summarized the Chinese organizational principle into a "different mode of association" (*chaxugeju*, 差序格局) (Fei, 1992: 71), which resembles ripples spreading from the centre of a body of water when a stone has been thrown into it. Social connections in China are non-equivalent and rely strongly on family ethics and kinship positions. Western social structures, in his opinion, have equivalent social connections and are constituted by autonomous groups and memberships. Fei's observation of Chinese social structure echoes the words of Max Weber. Weber noticed that behind that Chinese rational ethics were the principles of moral laws, which constitute a "complex of useful and particular traits" (Weber, 1964: 235). Chinese characteristic social patterning could also be attributed to the ethical system in Confucian teachings, in which the fundamental traits in personal relationships are summarised into five basic ethics (*lun* 伦) (Liang, 1987). The similar patterning of family to non-family relationships also appear in Chinese five *lun,* which constitutes the basic social order within Chinese society (King, 1985).

Recent scholars have considered *guanxi* culture from a micro and interpersonal perspective. Empirical researches taking this perspective proved the importance of kinship system in Chinese *guanxi* culture. Jacobs (1979: 237) defined *guanxi* as "particularistic ties", which consists of ascriptive ties, such as family members, or people from the same native place. Family and extended family ties are also the foundation and starting point of establishing other types of *guanxi* connections (Cohen, 1970; Baker, 1979). They sometime bridge between family ties with non-ascriptive ties and forms a sense of intimacy beyond the realm of immediate family. Jankowiak (2008) found the phenomenon where individuals address unfamiliar friends as family members such as sisters or uncles, in order to establish an intimate and useful connection.

Ethnographic research found that *guanxi* is often associated with sentiments (*renqing, ganqing*), face (*mianzi*), gifts (*liwu*) or favours; it is also deliberately maintained, cultivated and even manipulated to facilitate other social activities (Hwang, 1987; Yang, 1994; Yan, 1996; Kipnis, 1997; Smart, 1999). These studies on *guanxi* practice vividly describe the *guanxi* dynamism.

As a word frequently used to describe social connections, *guanxi* is often compared with social network in the Western academia. Some scholars considered *guanxi* different from social network (King, 1985; 1991), while empirical research on *guanxi* suggested that *guanxi* practices could not be fully interpreted using the framework of social network, and thus *guanxi* has certain features that make it essentially Chinese (Bian, 1994; Bian and Ang, 1997; Burt, 1992; Burt, 2002; Xiao and Tsui, 2007). Another group of scholars disagreed with the uniqueness and Chineseness of *guanxi* (Lin, 2001; Wellman, 2001; Wellman, Chen and Dong, 2002). They tried to incorporated *guanxi* culture within a larger conceptual framework of social network. Nevertheless, they still acknowledge the cultural influence of *guanxi* in Chinese societies.

**1.1.2 Functional Practices of *guanxi* in Business and Politics**

Studies conducted in Mainland China highlighted *guanxi* as a source of informal political power besides formal bureaucratic power, and in some cases even "defuse and subvert the elaborate regulations and restrictions" (Yang, 1994: 320). The administration "governed by human relationships" (Pye, 1992: 29) hampered Chinese government credibility in the Mainland, and has still exist in Chinese contemporary political system (Gold, 1986; Bian, 2002; Tsai, 2007).

Market economies in both Mainland China and in TW-HK-M have prospered since the mid-1990s. Researchers observed a pragmatic practice of business *guanxi* in both Chinese societies. Studies on Chinese family companies located in both Mainland and other Chinese communities have found that family businesses often demonstrated nepotism and paternalism characteristics (Cohen, 1970; Wong, 1985). It is sometimes used as a business strategy to provide informal support and protections, and is utilized to raise venture capital (Hamilton, 1998; Wank, 1999; Xin and Pearce, 1996; Luo, 1997; Hsing, 1998).

However, mainland Chinese Government plays a relatively strong role in the market economy. Study showed that business *guanxi* in the Mainland has been intentionally manipulated to establish connections with government officials in order to gain "back door access" in a state-dominated economy (Wank, 1999).

Turning to the 21st century, Chinese society in Mainland has moved towards a new model that demonstrates a combination of political socialism and economic capitalism. TW-HK-M, meanwhile, shares closer ties with the world economy. Under such circumstances, has *guanxi*, or social connections in China, co-evolved with other societal changes?

Some scholars see *guanxi* as the cultural consequences of the ancient Chinese social structure. *Guanxi* culture is thought to be challenged by two forces: one from political rationalization and one from the influence of market economy.

Political rationalisation and bureaucracy is believed to be an important factor accounting for the contemporary transformation of *guanxi* culture. Potter (2002: 183) argued that traditional Chinese *guanxi* will work as "a complement to rather than substitute for formal institutions" in the face of China's preceding legal reforms. Guthrie regarded *guanxi* as an "institutionally defined system" (Guthrie, 1998: 255) that could be shaped by the evolving political regime.

Chinese economic reform and the increasing influence of the market mechanisms are believed to be another driving force of the transformation of *guanxi*. For example, Yang speculated that abusive gift economy between government and market might gradually diminish with the market gradually replacing the state in resource distribution (Yang, 1994). Other researchers also documented the decline of cadre privilege (Walder, 1995) and the weakening of *guanxi* between officials and entrepreneurs (Ma and Cheng, 2010) with institutional changes in market-oriented reforms.

**1.2 Our Contribution**

As reviewed above, extensive research has contributed to the conceptualization of *guanxi*, and has examined the practice and transformation of *guanxi* culture. However, there remain two gaps in the literature: First, existing empirical research was mainly conducted during the early stages of Chinese reform in the 1990s, this was when China was still a largely socialist economy. As Chinese societies are undergoing dramatic social, economic and political changes, it is important to refine our understanding of *guanxi* in topics such as interpersonal relationships, informal politics and business network in contemporary Chinese society. Second, in many of the works on *guanxi,* the concept is often traced back to its Confucian root. However, these societies have followed divergent historical paths and developed different political and economic systems. Thus, it is important to be aware of the differences between Confucian societies inside and outside of Mainland China.

Based on the gaps identified in the literature, this research explores the following two questions:

**RQ 1**: Against the background of political rationalisation and economic reform, has the interpretation of *guanxi* changed in contemporary popular discourses?

**RQ 2**: When compared Mainland China to TW-HK-M, to what extent does *guanxi* adapt to different economic, social and political environments?

Unlike previous studies of *guanxi*, this research takes a unique approach to examining contemporary *guanxi* culture. Data come from everyday discourses on social media. Text data of *guanxi* discourses in this research were collected from Twitter, which is the only popular microblogging site that has users from both Mainland and the greater Chinese communities, providing an ideal public arena for a comparative analysis of two Chinese popular discourses.

This type of evidence of how users talk about *guanxi*, rather than how they respond to surveys or what can be observed by scholars, reveals a much more diversified interpretation of *guanxi* than inspecting it within a particular social activity or contextualising it within a specific community. As Kipnis argued (1996), analysing language surrounding *guanxi* can explore the richness of *guanxi* culture without over-generalising the concept beyond its meanings.

# 2. Methods

**2.1 Data**

    **2.1.1 Data Collection**

Data in this research were collected from Twitter's Search API[1]. The keywords in the search queries were simplified Chinese *guanxi* character ("关系") and traditional Chinese *guanxi* character ("關係"). Keyword search queries for recent tweets were sent every day over a period of three weeks from May 12th to June 1st, 2015. After combining the results and deleting duplicate tweets[2], there were 18,833 distinct tweets containing simplified Chinese *guanxi* and 9,572 distinct tweets containing traditional Chinese *guanxi*.

    **2.1.2 User Location**

This research used the different written forms of *guanxi* as proxies for two *guanxi* cultures. To justify this argument about written Chinese as indicative of Chinese societies in different regions, the two groups of *guanxi* tweets were compared by user locations. Location information (where available) was extracted from the metadata of all tweets, and was used for the comparison of the geographic distribution of users tweeting in simplified Chinese and in Traditional Chinese. In this research, user timezone is chosen as the main indicator of locations (see Appendix A for more details on the geographic information about the tweets).

---

[1] The returned result type was set to be "recent" tweets. According to Twitter's official documentation, returned results are likely to be incomplete. See: https://dev.twitter.com/rest/public/search
[2] Each tweet has a unique id ("id_str" in every status object), which was used as filter to delete repeated tweets in original dataset.

The number and proportion of tweets from each region are given in Table 1. 31.5% of tweets containing the Simplified Chinese character *guanxi* were sent by users located in Mainland China, but the percentage of users located in Mainland China among Traditional Chinese tweets was only 17.91%. A large percentage of users tweeting the Traditional Chinese *guanxi* were located in Taiwan, Hong Kong, and Macau. Therefore, the user language, either Simplified Chinese or Traditional Chinese, overlaps with the regions in which the users are located. Noticing that 2.24% of Simplified Chinese tweets were sent by users from Singapore and Malaysia, these tweets were deleted to enable comparison between Mainland China and TW-HK-M China. The final dataset contained 11,417 tweets about Simplified Chinese *guanxi* and 9,572 tweets about Traditional Chinese *guanxi*.

Table 1 Numbers and Proportions of Tweets in Different Regions

| Region [a] | Simplified Chinese | | Traditional Chinese | |
|---|---|---|---|---|
| | Number | Proportion | Number | Proportion |
| Mainland China | 3679 | 31.50% | 1212 | 17.91% |
| Taiwan, Hong Kong, and Macau | 767 | 6.57% | 2832 | 41.86% |
| East Asia | 1951 | 16.71% | 1173 | 17.34% |
| Overseas | 5020 | 42.98% | 1549 | 22.89% |
| Singapore and Malaysia [b, c] | 262 | 2.24% | N/A | N/A |
| Total | 11679 | 100.00% | 6766 | 100.00% |

*NOTE.* The total number of tweets in this table is tweets with user timezone information. [a] User timezones were manually coded into five categories: Mainland China, Taiwan, Hong Kong, and Macau, East Asia, Overseas (excluding East Asia), Singapore and Malaysia. [b] After 1976, Singapore adopted the same system of Simplified Chinese characters as that of the People's Republic of China. [c] Malaysia adopted Simplified Chinese characters in 1981.

**2.1.3 Building Corpora**

The two corpora in this research were organized and cleaned in three steps: punctuation filtering, word segmentation, Part-of-Speech tagging and filtering. In the first step, raw texts in Simplified Chinese and in Traditional Chinese were aggregated separately by extracting tweet text from the dataset, and were processed to filter punctuations, emoji and other symbols. Then a Python module for Chinese word segmentation and part-of-speech tagging, Jieba[3], was utilised to segment words in each sentence in both corpora.[4,5] After tokenising, the corpora consist only space-spliced words. Not all words are necessary for understanding the context of sentences (Jurafsky and Martin, 2014). By denoting all words in both corpora with their according part-of-speech, we only keep words that are semantically meaningful, which includes nouns, verbs, adjectives and adverbs.

**2.2 Semantic Network Analysis**

**2.2.1 Word Co-occurrence**

In understanding word senses, one group of linguists suggested approaching word meaning via its relationship with other words in the same sentence, taking a structuralism perspective (Firth, 1957; Saussure, 2011; Evans, 2005). To structuralise language in concrete forms, semantic network was proposed by applied linguists in the field of computer science (Quillian, 1963; Collins and Quillian, 1969).

As with other types of networks, semantic networks consist of vertexes and edges. Vertexes, or nodes, are concepts; edges can either be constructed by existing paradigmatic knowledge of linguistic relationships such as thesaurus (Fellbaum, 1998; Kozima and Furugori, 1993), or by the totality of structural relationships, such as co-occurrence (Freeman and Barnett, 1994; Danowski, 1993; Doerfel, 1998).

In this research, we used word co-occurrence to construct the semantic network. As defined by Freeman and Barnett (1994), the links in a semantic network are the co-presence frequency of a pair of concepts in a given slicing window of the text. The slicing window in this research was set to be three words[6], and WORDij[7] was used to count the co-occurrence of word pairs and frequency of words.

---

[3] GitHub page of Jieba. https://github.com/fxsjy/jieba
[4] An alternative Chinese language segmentation software is NLPIR-ICTCLAS (http://www.nlpir.org/?action-viewnews-itemid-323). The two modules have similar Chinese word segmentation functions but Jieba has a dictionary that has better support for traditional Chinese characters (https://github.com/fxsjy/jieba/raw/master/extra_dict/dict.txt.big).
[5] Chinese natural language is written in a sentence without space, and thus splitting Chinese sentences is based on existing knowledge (pre-installed lexicon) of the common expressions.
[6] For the aim of conceptualizing *guanxi*, a smaller slicing window for identifying fixed expressions is preferred. (Church and Hanks, 1990).
[7] WORDij: http://www.content-analysis.de/2010/09/24/wordij.html

### 2.2.2 Measuring Associations

The association of each word pair $x$ and $y$ can be measured by their co-occurrence frequency. However, words in natural language are distributed following the power law (Jurafsky and Martin, 2014) (see Appendix B for information on word frequency distribution of corpora in this research). Highly associated word pairs denoted by co-occurrence frequency, therefore, are likely to co-locate due to chances rather than semantically meaningful connections. For this reason, it is important to use a different measurement, one that can measure the likelihood that the target word pair $(x, y)$ co-occurred by chance irrespective of the frequency of the words. One such measurement using the ratio of observed occurrence and the expected probability of occurrence if two words appeared independently was proposed by Church and Hanks (1989), and is called Pointwise Mutual Information ($PMI$):

$$PMI(x, y) = log_2 \frac{P(x,y)}{P(x) \times P(y)} \qquad (1)$$

Where $P(x)$ and $P(y)$ are words' counts in the corpus, normalized by $N$, the size of the corpus. $P(x) \times P(y)$ measures the likelihood that $x$ and $y$ would co-occur in the same corpus if they were independent. $P(x, y)$ is the joint probability of $x$ and $y$ observed to be co-located in a window of $w$ consecutive words.

Using $PMI$ as a measurement of words association takes into account the scenarios, in which two words only locate within the slicing window by chance. However, it has two drawbacks as a co-occurrence statistic: First, the denominator $P(x) \times P(y)$ could be extremely small for very rare words, making low frequency word combination scores disproportionately high, which would interfere with detecting more meaningful and commonly used word pairs. Second, there is a lack of an upper bound for PMI value, making it practically difficult to define a high PMI score. To solve these two problems, Bouma (2009) devised a method to normalise PMI, which controls the influence of rare words and defines an upper boundary. The normalised measure, nPMI, is defined as:

$$nPMI(x, y) = \frac{PMI\ (x,y)}{-log_2\ (P(x,y))} \qquad (2)$$

In this research, the $nPMI$ scores of all the word pairs were used as the weight of the edge between the two nodes in the semantic network.

**2.2.3 Visualising Semantic Network**

After calculating the $nPMI$ scores for all the word pairs in the Simplified Chinese and Traditional Chinese corpora, edges (weighted by $nPMI$ score) and nodes (distinct words) can be visualised in two semantic networks: one in Simplified Chinese and the other in Traditional Chinese. Both networks were further filtered to exclude rare words and edges with lower $nPMI$ values.

First, we filtered out words with very small frequencies (see Appendix B); words that appeared three times or less in the Simplified Chinese and only once in the Traditional Chinese corpus.

Next, when choosing a threshold for significant word pairs, we used a percentile threshold for determining the most relevant word associations[8]. The major consideration in choosing a threshold is not whether the $nPMI$ value is statistically significant, but whether it is practically important for interpretations. Based on qualitative comparison of different percentile thresholds, we applied a threshold of the 30$^{th}$ percentile of sorted $nPMI$ scores, which was 0.1955 for the Simplified Chinese corpus and 0.2546 for the Traditional Chinese corpus (See Appendix C for the justifications of choosing the percentile threshold).

Two semantic networks consisting of edges with weights ($nPMI$ scores) exceeding the threshold values could now be constructed based on the filtered $nPMI$ scores. After removing rare words and filtering unimportant word pairs, there were 20,710 words and 134,072 edges in the Simplified Chinese semantic network and 12,137 words and 119,461 edges in the Traditional Chinese semantic network.

The semantic network of *guanxi* in Simplified Chinese and in Traditional Chinese were displayed separately, each node represents a word from *guanxi* corpora. Node size indicating the node strength, which is the sum of weights attached to edges belonging to a node (Barrat et al., 2004). Edges are undirected and weighted. Two semantic networks were visualised using network analysis tool: Gephi (Bastian, Heymann and Jacomy, 2009), displayed by applying Force Atlas 2 algorithm (Jacomy, Heymann, Venturini, and Bastian, 2014). Word pairs connected by edges with higher association values ($nPMI$ scores) are attracted proportionally closer than others.

---

[8] Another method to identify significant edges is to shuffle the nodes (words), shuffle the network while remain the node degrees, and rewire nodes randomly with each other. By calculating t-value, a p-value can be found using Student's t-distribution. If p-value is less than the conventional threshold for statistical significance, 0.05, then the null hypothesis (two words are independent) is rejected and the alternative hypothesis (they are associated semantically) is accepted (see Newman, Watts and Strogatz, 2002).

Semantic networks contain a large amount of concepts and concept relations. To categorise concepts, community detections were applied on both *guanxi* semantic networks using Modularity partition algorithm in Gephi (Newman, 2006). After partitioning, the Simplified Chinese semantic network of *guanxi* was divided into 43 communities, with Modularity score equals to 0.431; the Traditional semantic network was partitioned into 24 communities, with a Modularity score of 0.339[9]. Themes of each sub-community were summarized by the central concepts, which were defined as the top 5 percent of nodes ranked by their node strength in descending order (see Appendix D for details on the justification of choosing the 5 percent threshold). Figure 3 and Figure 4 shows the clustered networks of Simplified Chinese *guanxi* and Traditional Chinese *guanxi*. Figure 2 summarizes the methodology of this research using a flow chart.

---

[9] During the process of modularity optimization, both modularity scores were highest when resolutions were 1.0. Modularity scores were used to compare between possible partition results in the same network, not for comparing between two networks.

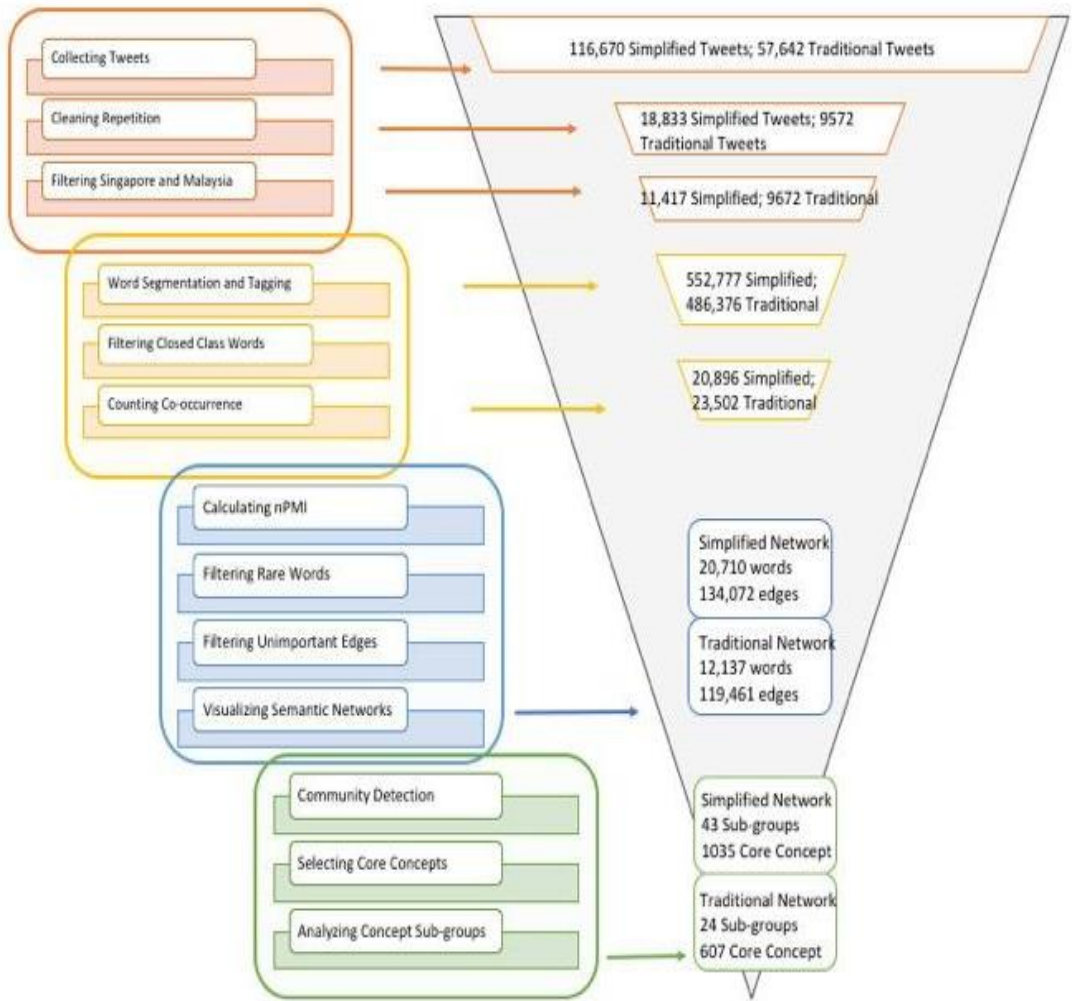

Figure 1 Methodology and Data Cleaning

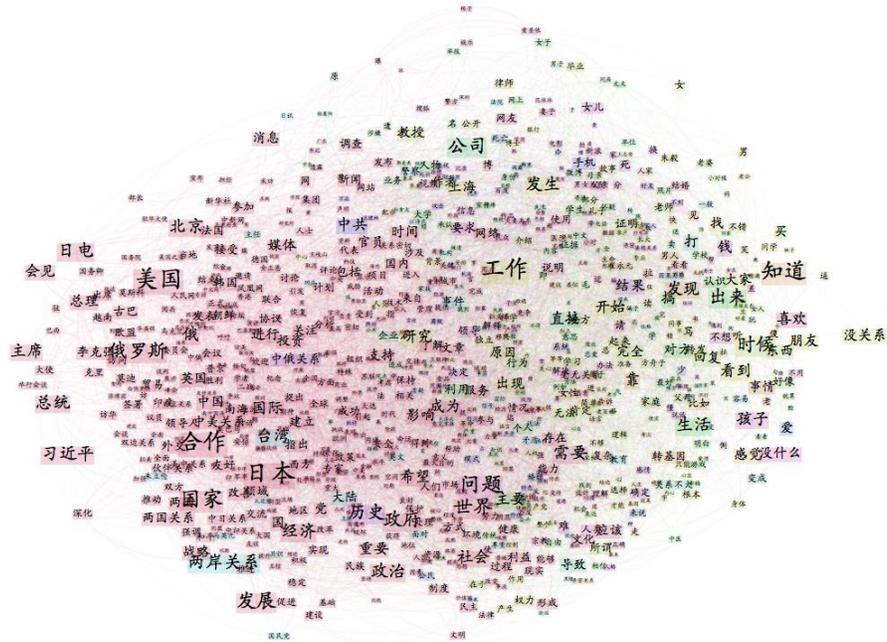

Figure 2 Semantic Network of Simplified Chinese *Guanxi* (N=20710. Only central concepts were displayed. Label box and edge colours indicate concept communities)

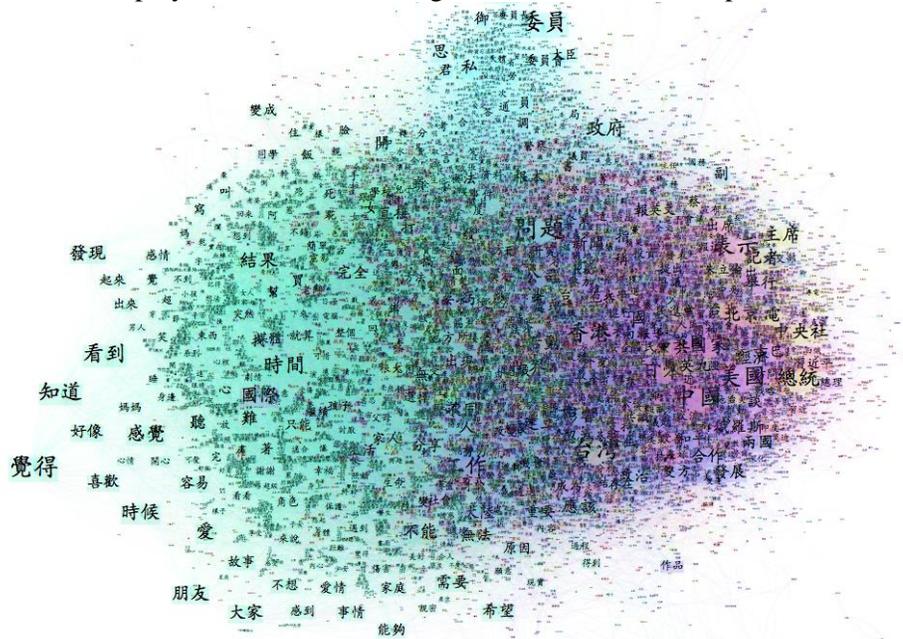

Figure 2 Semantic Network of Traditional Chinese *Guanxi* (N=12137. Only central concepts were displayed. Label box and edge colours indicate concept communities)

## 3. Results

By comparing the concept sub-communities of the two *guanxi* semantic networks, we found that although the two discourse networks share similar themes, including interpersonal relationships, business-political networks, and state relations. However, they demonstrated different interpretations within these topics.

**3.1 *Guanxi* in Interpersonal Relationships**

As explained previously, *guanxi* consists of social norms and traditions relating to interpersonal connections. Through examining the Simplified and Traditional *guanxi* semantic networks, it can be seen that they both have one or several sub-groups concerning interpersonal relations. However, *guanxi* in the domain of interpersonal relationships have demonstrated different focuses in the two Chinese societies.

**3.1.1 Mainland: Intimacy and Traditions**

There are two concept communities in Simplified Chinese semantic network related to interpersonal relationships. The first one ranks third largest sub-community (see Figure E1 in Appendix E), highlighting intimate *guanxi*; the second one is the fourth largest sub-community (see Figure E2 in Appendix E) in the Simplified Chinese network, with a focus on Chinese traditions.

Familism in traditional Chinese culture consists of members within an extended family. Particularly in rural China, the boundary of a family could be as broad as a lineage or a clan (King, 1991). However, as shown in Figure E1, the highest ranked concept community in family *guanxi* in Simplified Chinese is more concerned with the nuclear family than it is with lineage. Concepts related to romance are emphasised by relationships such as "Marriage relations" (夫妻关系), "Lovers" (情人) and roles such as "Wife" (妻子), and "Girlfriend" (女友). The extended family retains its influence on individual through interfering with marriage. For instance, remotely or closely related "Aunts" ("七大

姑八大姨") are common matchmakers who are actively involved in seeking and assessing an ideal match for the "Boy" ("男孩") or "Girl" ("女孩").

Concepts of *guanxi* as a tradition or cultural heritage belong to the fourth largest concept sub-community in the Simplified Chinese semantic network (displayed in Figure E2). In this community, feelings and social consequences of *guanxi* tradition were discussed. Originating from "*lun*" in Confucianism, the *guanxi* tradition is characterized by "Patriarchy" ("家长制"), highlighting the hierarchical relationship between "Old generation" ("老人") and "Children" ("孩子") within the "household" ("家里"). In addition, the traditions are connected through attributive ties, especially by "Consanguinity" ("血缘关系"), which is the sharing of bloodlines.

There is, however, a bittersweet emotion regarding traditional *guanxi* that is embedded in Chinese family morals. Centred on the deeply rooted *guanxi* traditions, there is nostalgia for a harmonious household and concern about overall social equality. On one hand, nostalgia for the *guanxi* tradition is linked to harmonious family relationships and cohesive family structure. For example, emotions like "Happiness" ("幸福") and a sense of "Inheritance" ("传承") are all inherent characteristics of the *guanxi* tradition. On the other hand, after Mao's anti-tradition social movement from 1966 to 1976, the "Cultural Revolution" ("文革") had challenged traditional values in urban and rural China tremendously, which is also reflected in this concept community via concepts such as "Revolution" ("革命"). With the Chinese modernization process, there are also negative attitudes toward traditional *guanxi* norms. One criticism of *guanxi* tradition is the over-emphasis on one's family "background" ("背景"). If used illegitimately to obtain private profits, *guanxi* might pose a threat to social "justice" ("公平"). It might also jeopardize the legal rights and equal opportunities of other individuals from less privileged families.

**3.1.2 Taiwan, Hong Kong, and Macau: New Wine in Old Bottles**

Concepts regarding interpersonal relationships form the largest concept communities in the Traditional Chinese semantic network. This ranking order suggests discourses on *guanxi* are more often associated with social connections in TW-HK-M than they are in Mainland. The two Chinese communities also have varying emphasis on interpersonal relationships. As analysed in the Simplified Chinese network, intimate *guanxi* and *guanxi* heritage are heated topics in the domain of relationships. In the Traditional Chinese *guanxi* network, however, central concepts are more general descriptions of social networks, without emphasizing intimate *guanxi* or debating on the traditions of Chinese family morals (see Figure E3 in Appendix E).

Chinese users in TW-HK-M show a form of interpersonal *guanxi* that is closer to the modern concept of social connections in many Western societies. traditional *guanxi* relations and the traditional moral requirements for the family are both missing form this sub-community.

This concept sub-group in Traditional Chinese embraces a wide range of issues concerning interpersonal relations ranging from types of social relationships such as "Family" relations ("家庭"), "Schoolmates" ("同學"); to "Social roles" ("社會角色") of individuals in these relations, including "Parents" ("父母"), "Wife" ("妻子"); and to emotions and feelings about relationships, for instance, "Love" ("爱情"), "Trust" ("信任"), "Intimacy" ("親密"), or "Dislike" ("討厭"). Although the types of *guanxi* in this community are diverse, which include weak ties such as "Friends" ("朋友") and "Colleagues" ("同事"), there is still more mentions of members in a nuclear family.

**3.2 *Guanxi* in Two Market Economies**

As reviewed earlier, much research on the Mainland China and TW-HK-M markets has noticed the pervasiveness of *guanxi* practice in Chinese companies. Networks in this research also prove that *guanxi* in economic activities is an important topic in both societies, but business *guanxi* functions differently in the two contexts.

**3.2.1 Mainland China: *Guanxi* and Rent-seeking**

In one concept community of the Simplified semantic network, *guanxi* is frequently associated with rent-seeking behaviors, through manipulating *guanxi* ties with government officials. Concepts in this domain cover both the major players in rent-seeking processes and the corresponding social consequences of such practices. This concept sub-community on business *guanxi* in Simplified Chinese is displayed in Figure E4.

Unlike what was observed in the early stages of Chinese economic reform (Yang, 1994) when *guanxi* was mainly used to acquire priority in the distribution of raw materials for production. In the contemporary Chinese economy, the abuse of power and position extends to the "Capital" ("资本") and "Investing" ("投资") process. Industries referred to in this concept community ranged from primary or infrastructural industries such as "Agriculture" ("农业") and "Railway" ("铁路"), to non-traditional or emerging industries such as "Tele-communication" ("通信"), "Technology" ("科技"), and "Finance" ("金融"). Regarding the soil in which the rampant corruption is embedded, the "Cliques" ("集团") between "Local" ("地方") officials and "Corporations" ("企业") are all mentioned. "State-owned companies" ("国企") were particularly highlighted under this topic.

Besides, there are also discussions on regulating and restricting *guanxi* practice in business. The violations of formal "Law" ("法") and public "Moral" ("道德") have provoked "Reflection" ("检讨") on the Chinese "Bureaucratic" system ("制度") in order to ensure "Equality" ("平等") in market competitions, providing equal opportunities for incorrupt competitors.

### 3.2.2 Taiwan, Hong Kong, and Macau: Business Network and Regulated *Guanxi*

In traditional *guanxi* semantic network, topics on business *guanxi* locate in separated community with political *guanxi*. Such a separation is different what is observed in simplified *guanxi* semantic network of Mainland users.

The second largest concept community in the Traditional Chinese semantic network consists of concepts related to business *guanxi* (see Figure E5). Unlike Simplified Chinese business *guanxi*, this concept community tends to focus on *guanxi* in economic activities as a business strategy, which separates the market from government officials. "Enterprises" ("企業") and "Companies" （"公司"） are major players in the business *guanxi* network. *Guanxi* in their business "Strategy" ("戰略") is considered to be access to personal "Communications" （"溝通"） and as "*Guanxi* Capitals" ("人脈") for "Opportunities" ("機會") and "Investment"("投資"). Compared to enterprises in Mainland, who have to invest in establishing *guanxi* resources with government officials; *guanxi* in TW-HK-M is a form of human resource and an informal business strategy brought by and maintained through personal ties between business partners.

Concepts of political *guanxi* are located in a different concept community in the Traditional Chinese semantic network, ranking as the fourth largest concept community (see Figure E6). In this concept community, political *guanxi* is confined to the legislative, executive and administrative divisions. Individuals who have political power are mentioned in a same political network, consisting of "Civil servant" ("公務員"), "Committee member" ("委員") and "Member of Parliament" ("議員"). This concept community also demonstrates a clear separation between the "Private" ("私") domain and the "Public" ("公") domain, as well as highlighting "Regulation"（"制度"）, and "Law"（"法"） as constraints of the political *guanxi*.

Therefore, business *guanxi* in TW-HK-M is linked more closely to networks among personnel and less with governments. Political *guanxi* also raises alarms in these societies, but is restrained by written constitutions and formal laws.

**3.3 *Guanxi* in Power Relations: An Extended Context**

Most scholarly writings on *guanxi* considered the concept from micro- and meso-levels, focusing on the role of *guanxi* in networking activities between individuals or organisations. However, community-clustering on both *guanxi* semantic networks captured a new context of using *guanxi*, which is in power dynamics between states.

Figure E7 and Figure E8 display the largest sub-community of the Simplified Chinese *guanxi* semantic network and the third largest sub-community of the Traditional Chinese *guanxi* semantic network; both communities contain concepts describing relationships between states.

In Simplified Chinese, most of the concepts of *guanxi* between states concerning diplomatic taxonomies and country names. Interestingly enough, it also demonstrates many diplomatic strategies borrowed from Chinese interpersonal social norms. For example, sentiments such as "Family" (家庭), "Trust" (信赖), and "Responsibility" (责任) are used in the domain of international relations.

This finding of state relations in both discourses has two implications: First, it suggests that Chinese *guanxi* culture is also reflected in the state's foreign policies. This expands the realm of *guanxi* study to the state level, and to Chinese diplomatic relations with other countries. As highlighted by the personalized sentiments in this concept community, the promotion of Confucian values plays an important role in Chinese diplomatic strategy. Second, the fact that over 16% of all concepts in Simplified Chinese *guanxi* semantics and 11% in Traditional Chinese *guanxi* semantics reside within this state relation community also indicates that in both Mainland China and in TW-HK-M China, Twitter is used regularly as news following and sharing services. This is also proved by the appearance of newspaper and agency names such as "People's Daily" (人民网), "VOA" (美国之音) in the two communities.

## 4. Discussion and Conclusion

This research uses *guanxi* discourses from social media to explore the transformation of *guanxi* culture in contemporary popular discourses. Table 2 provides an overview of the concept communities analysed in this research. The largest four concept communities in each semantic network cover three major topics concerning interpersonal relationships, business *guanxi*, and state relations.

Table 2 Overview of Concept Communities

| Rank | Simplified Chinese *Guanxi* Semantic Network | | | | | Traditional Chinese *Guanxi* Semantic Network | | | | |
|---|---|---|---|---|---|---|---|---|---|---|
| | Topic | Nodes | | Edges Number | Central Nodes Number[a] | Topic | Nodes | | Edges Number | Central Nodes Number[a] |
| | | Number | Proportion | | | | Number | Proportion | | |
| 1 | States Relation | 3528 | 16.97% | 19633 | 176 | Interpersonal *Guanxi* | 3224 | 26.61% | 20968 | 161 |
| 2nd | Business-Political *Guanxi* | 987 | 4.75% | 2607 | 45 | Business *Guanxi* | 1444 | 11.89% | 6727 | 72 |
| 3rd | Interpersonal *Guanxi*-Intimate | 815 | 3.94% | 1632 | 41 | States Relation | 1358 | 11.18% | 5686 | 68 |
| 4th | Interpersonal *Guanxi*-Tradition | 667 | 3.21% | 1427 | 34 | Political *Guanxi* | 1313 | 10.81% | 7358 | 66 |

NOTE. Total concept number in Simplified Chinese *guanxi* semantic network is 20710. Total concept number in Simplified Chinese *guanxi* semantic network is 12137 [a] Central Topics are defined as the top 5% of concepts ranking by node strength.
[a] Central Topics are defined as the top 5% of concepts ranking by node strength.

**5.1 Transformation of Family Values**

We found that in terms of interpersonal relationships, both Mainland and TW-HK-M China have demonstrated an emphasis concerning the nuclear family. However, concept communities on interpersonal relationships show that the significance of traditional lineage families and their associated morals are declining in both the Mainland and TW-HK-M.

Despite the transformation of the family structure reflected in *guanxi* discourses, in the Simplified Chinese *guanxi* semantic network, special attention is still paid on the traditional values and morals in Mainland China. In the discussion of traditions, many concepts reflected negative attitudes towards the traditional system of hierarchy within the family and towards the over-emphasis of family background.

This widespread doubt about traditional family *guanxi* within the Simplified Chinese semantic network will be likely to result in a clash with official ideology. The transformation of the value system in official discourses aims at addressing the moral vacuum in contemporary Chinese society after the fall of Marxism ethics in face of the market economy (Bell, 2010). However, we can observe from the semantics of the *guanxi* tradition online that there is a mismatch between the popular discourses and the official ideology. This might lead to a sense of uncertainty and confusion about the ongoing transformation of ideology or doctrines in the Mainland.

### 5.2 *Guanxi* in Two Markets

Our research also showed in the Simplified Chinese *guanxi* semantic network, the concept *guanxi* bridges industries and the government, forming cliques in which members benefit each other through the exchange of political and economic resources. Such a phenomenon is also often the target of criticisms in the Simplified Chinese discourses. Meanwhile, business *guanxi* in the Traditional Chinese semantic network is separated from the political *guanxi* concept community. This clear division of politics and business ensures a fair environment for the market.

Whether or not political *guanxi* is intertwined with the business networks, and how political leverage is constrained to avoid corruption and rent-seeking behaviours mirror institutional differences in two Chinese societies.

The close-knit connection between political power and economic entities in the Mainland can be attributed to the lack of constraints over bureaucrats in Chinese state capitalism. Since the 1980s, the Chinese Communist Party has promoted a market economy featuring strong government control. With the market growth and capital accumulation, the lack of effective structural constraints on officials leads to rampant corruptions and rent-seeking behaviours. The lack of efforts for political reform might lead to the continuing spread of corruption and rising dissatisfaction concerning the party's legitimacy. Eventually, this will result in criticism of and distrust in the government (Fukuyama, 2014). By contrast, in TW-HK-M, a globalized market and relatively effective constraints on bureaucratic power (Jones, 1994) help to prevent officials from using *guanxi* to obtain illegal profits. *Guanxi* connections in these market economies are more difficult to be manipulated by individuals.

A regulated market economy requires clear laws formalizing the types of *guanxi* practices that are acceptable and those that are forbidden. This is not only important for the Chinese market economy, but also significant for Chinese political legitimacy in the long run. Business *guanxi* in Taiwan, Hong Kong, and Macau demonstrates the possible variance of *guanxi* culture in a regulated market economy. For leaders in the Mainland, these economies might provide solutions for tackling rampant corruptions.

**5.3 Limitations**

Whilst semantic networks generated from popular discourse could reflect the understanding of *guanxi* among Chinese Twitter users, there are still some limitations of the study.

First, Twitter is unfamiliar to most of Internet users in Mainland China due to the government block (the Great Firewall); in particular, it has not been used by people who live in rural China or who do not have sufficient technological skills to bypass the Internet block. This makes Mainland Twitter users a relatively special group of Chinese Internet users and also lacks representativeness of the general Chinese population. However, there is an advantage in using discourses of Mainland Twitter users to study *guanxi*. This group tends to be very sensitive to Chinese politics and is less constrained by government censorship, which allows for the observation of critical opinions concerning Chinese political *guanxi*. Second, as the relationship between language and mind is complex, popular discourses on *guanxi* could not contribute more details regarding the dynamic process of practicing *guanxi* in offline settings. Third, concepts represented by segmented words, only provide a limited window into understanding what people think about or believe in, and thus certain context of the concepts is lost when processing language.

To conclude, this research revisited *guanxi* culture in different Chinese societies by analyzing its linguistic representations in popular discourses. The concept g*uanxi* bridges individuals and societies in the Chinese context. Reifications of the concept *guanxi* had not only provided understanding of *guanxi* in contemporary Chinese societies, but had also served as an interpretive device for studying the political and economic systems in which *guanxi* cultures are embedded.

**Declaration of conflicting interest**

The author(s) declared no potential conflicts of interest with respect to the research, authorship, and/or publication of this article.

**Funding**

The author(s) received no financial support for the research, authorship, and/or publication of this article.

# Appendices

**Appendix A:**

**Tweet Country, User Timezone and Description of User Location**

Three types of location information were extracted from metadata of tweets: tweet country, user timezone and description of user location. Table A1 details features of each type.

Table A1 Comparison of Three Types of Location Information

| Type | Location [a] | Editable | Missing Level | Simplified (N=18,833) | Traditional (N=9572) |
|---|---|---|---|---|---|
| Tweet Country | Tweet entities | NO | High | 98.46% | 98.47% |
| User Timezone | User profile | NO | Low | 36.52% | 29.31% |
| Description of User Location | User profile | YES | Medium | 57.78% | 45.98% |

*NOTE.* [a] Location in the meta data of tweet or user profile.

Figure A1 and Figure A2 show the percentage of different regions of Simplified Chinese tweets and Traditional Chinese tweets, based on the tweet location entries (less than 2 % of total tweets on *guanxi*). The amount of geotagged tweets is relatively small compared to tweets with user timezone or description of user location. The distribution of regions corresponds with Table 2 and Figure 2.

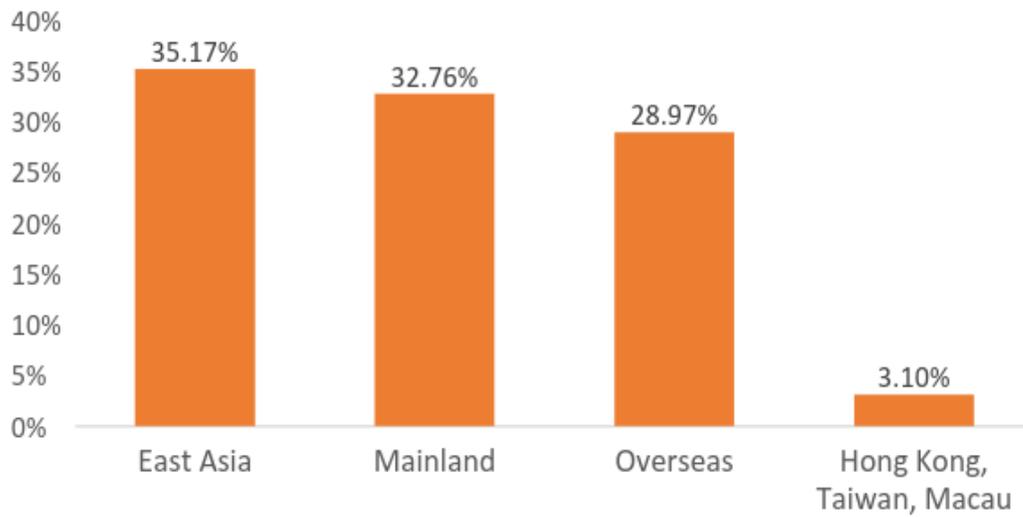

Figure A 1 Distribution of Simplified Chinese Tweets' Locations by Region, based on Tweet Country (in Percentage)

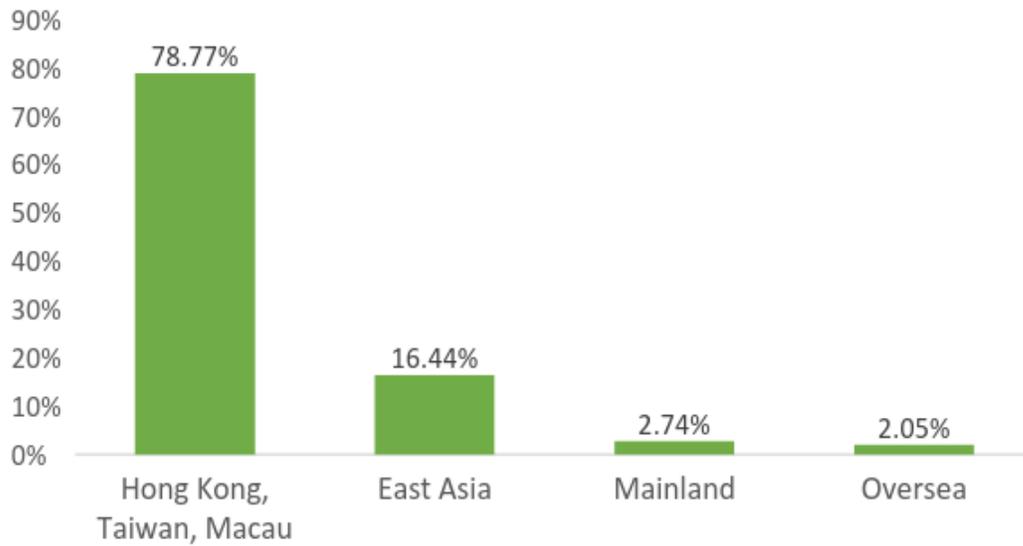

Figure A 2 Distribution of Traditional Chinese Tweets' Locations by Region, based on Tweet Country (in Percentage)

While tweet country and user timezone could be selected from a list of locations by users, description of location is more personalized. When analysing text in these location descriptions, different characteristics emerged between Simplified Chinese users and Traditional Chinese users: Words related to political values were used by Simplified Chinese users in their location profile. For instance, democracy (*minzhu*, 民主), justice (*zhengyi*, 正义), censorship (*hexie*, 河蟹, translated literaly river crab, but is used as an ironic reference to government censorship) appeared in user location descriptions. In comparison, although Traditional Chinese users also indicated creative usage of location names such as "*dahe* state" (*dahe* is the translation of Japanese word やまと, referring to the spirit of Japanese), they rarely use politically related terms in the location profile.

**Appendix B:**
**Measurements of Co-occurrence**

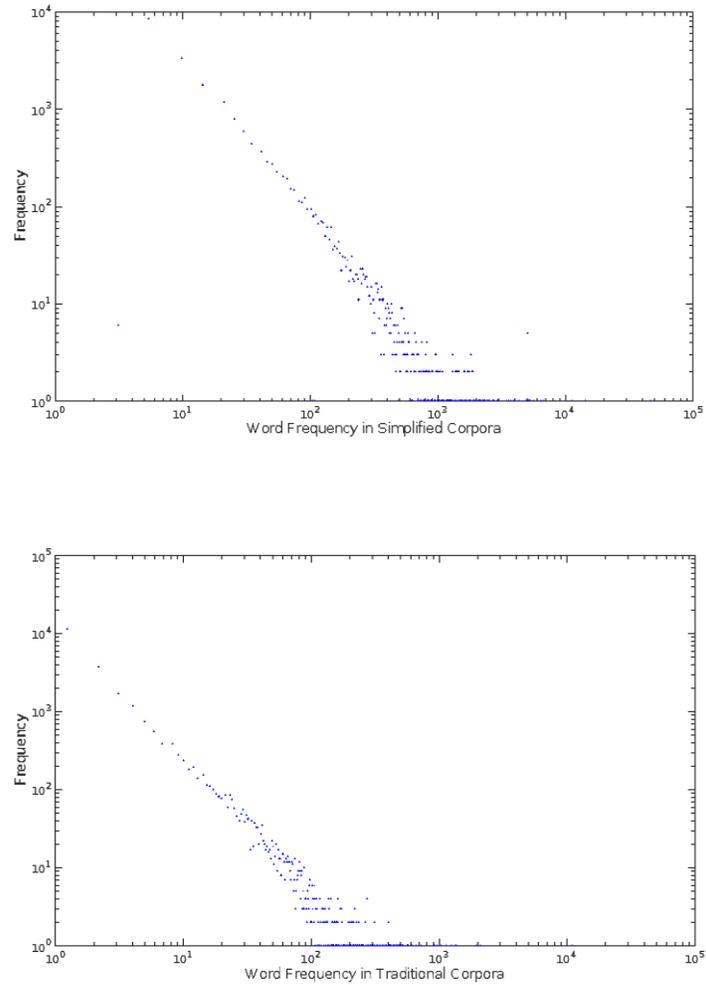

Figure B1 Log-log Plot of Word Frequency Distributions. (a) Histogram of word frequency in Simplified Chinese corpus on logarithmic scales. (b) Histogram of word frequency in Traditional Chinese corpus on logarithmic scales. Word frequency in both corpora follows the power law. *NOTE.* Distributions in these graphs were generated by taking binning centers of the original distribution of word frequencies.

# Appendix C:
# Choosing Percentile Threshold for nPMI (Edge Weights)

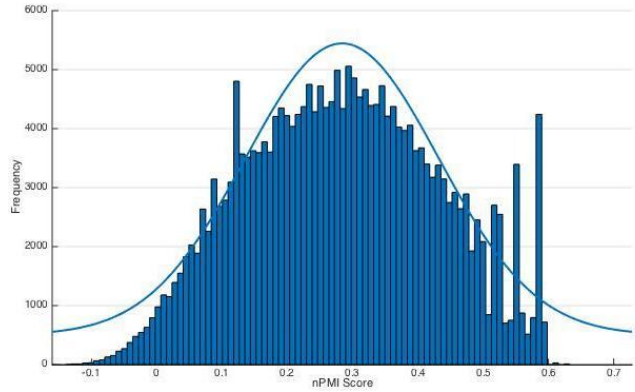

| Percentile | nPMI |
|---|---|
| 15% | 0.1218 |
| 30% | 0.1955 |
| 45% | 0.2621 |
| 60% | 0.3241 |

*N=225,789*

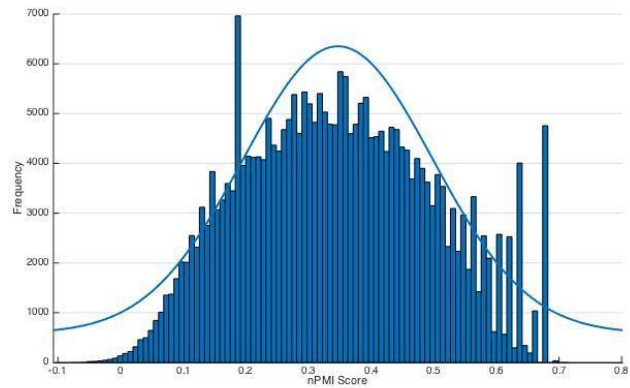

| Percentile | nPMI |
|---|---|
| 15% | 0.1819 |
| 30% | 0.2546 |
| 45% | 0.3200 |
| 60% | 0.3847 |

*N=268,862*

Figure C1 Histogram and Percentiles of nPMI Scores. (a) Histogram of word pairs nPMI scores from the Simplified Chinese corpora with superimposed best-fitting curves. Best-fitting curves were generated using Matlab function "normfit", which estimates the normal parameters: the mean μ and the standard deviation σ (b) Histogram of word pairs nPMI scores from the Traditional Chinese corpora with superimposed best-fitting curves. The table beside each histogram lists the 15th, 30th, 45th and 60th percentile value.

Table C1 nPMI Scores of Concepts Associated with "Family" in Simplified Chinese Corpus

| Stem Concept | Associated Concept | Chinese Pronunciation | 简体中文 |
|---|---|---|---|
| Family | Family Member | *Jiating Chengyuan* | 0.418922951 |
| Family | Harmony | *Hexie* | 0.406543334 |
| Family | Sibling | *Shouzu* | 0.382903188 |
| Family | Responsible for | *Chengdan* | 0.382903188 |
| Family | Only Child | *Dusheng Zinv* | 0.372256159 |
| Family | Decedents | *Ersun Houdai* | 0.36566327 |
| | 60th Percentile | | |
| Family | Bond | *Niudai* | 0.307447872 |
| Family | Mother | *Muqin* | 0.305804511 |
| Family | Grandparents | *Yeye Nainai* | 0.294195059 |
| Family | Power | *Tequan* | 0.288353023 |
| Family | House | *Fangzi* | 0.286935955 |
| Family | Money | *Qian* | 0.269709224 |
| | 45th Percentile | | |
| Family | Couple | *Fuqi* | 0.220490986 |
| Family | Children | *Er'tong* | 0.208665012 |
| Family | Conflicts | *Jiufen* | 0.200283635 |
| Family | Enhance(heritage) | *Hongyang* | 0.200283635 |
| | 30th Percentile | | |
| Family | Legal | *Hefa* | 0.186824383 |
| Family | Business secrets | *Shangye Jimi* | 0.176748222 |
| Family | Intensive | *Jinzhang* | 0.161316158 |
| Family | Migration | *Yimin* | 0.129775235 |
| | 15th Percentile | | |
| Family | Matter | *Shiwu* | 0.116537903 |
| Family | Background | *Beijing* | 0.095474473 |
| Family | Direct | *Zhijie* | 0.079280363 |
| Family | Feeling | *Ganjue* | 0.068154144 |
| Family | Beginning | *Kaishi* | 0.038173513 |

*NOTE.* Dashed Line breaks indicate four percentile thresholds

| Stem Concept | Associated Concept | Chinese Pronunciation | 中文词汇 |
|---|---|---|---|
| Family | Happiness | *Fulu* | 0.497581539 |
| Family | Harmony | *hexie* | 0.462837084 |
| Family | Women | *Funv* | 0.427650813 |
| Family | Conflict | *Moca* | 0.426223795 |
| Family | Status | *Chushen* | 0.38467518 |
| | 60th Percentile | | |
| Family | Network | *Renmai* | 0.381912965 |
| Family | Brother | *Didi* | 0.352997216 |
| Family | Children | *Zinv* | 0.33992685 |
| Family | Girlfriend | *Nv Pengyou* | 0.336175895 |
| Family | Patriarchy | *Jiazhang* | 0.327351797 |
| | 45th Percentile | | |
| Family | Tradition | *Guannian* | 0.319792942 |
| Family | Care | *Zhaogu* | 0.31710459 |
| Family | Couple | *Fuqi* | 0.306690948 |
| Family | Parents | *Fumu* | 0.305518742 |
| Family | Blood ties | *Xueyuan* | 0.276568314 |
| Family | Mother | *Mama* | 0.25726405 |
| | 30th Percentile | | |
| Family | Work | *Gongzuo* | 0.234205349 |
| Family | Talent | *Caineng* | 0.233389377 |
| Family | Security | *Anquan* | 0.229957355 |
| Family | Society | *Shehui* | 0.189722684 |
| | 15th Percentile | | |
| Family | Eating | *Chi* | 0.177696264 |
| Family | Hong Kong | *Xianggang* | 0.142291802 |
| Family | Guanxi | *Guanxi* | 0.168134512 |
| Family | Taiiwan | *Taiwan* | 0.100555165 |

*NOTE.* Dashed Line breaks indicate four percentile thresholds

Table C 2 nPMI Scores of Concepts Associated with "Family" in Traditional Chinese Corpus

As indicated in Figure C1, which shows the distributions of *nPMI* scores of Simplified Chinese and Traditional Chinese word pairs, both groups of word pairs are distributed symmetrically around their means; the mean score of *nPMI* of Traditional Chinese word pairs is slightly higher than scores for Simplified Chinese. Using percentiles avoids setting absolute values for different datasets. we justified the choice of a particular percentile threshold by looking at the table of *nPMI* scores for one word "family", with its associated words in both corpora (see Appendix C, Table C1 and Table C2). For both Simplified Chinese and Traditional Chinese concepts of "family" (*Jiating*, 家庭), associated concepts that score higher than the 30[th] percentile value of *nPMI* are more closely related to the definition of family. Thus, we applied a threshold of the 30[th] percentile of sorted *nPMI* scores, which was 0.1955 for the Simplified Chinese corpus and 0.2546 for the Traditional Chinese corpus. Two semantic networks consisting of edges with weights (*nPMI* scores) exceeding the threshold values could now be constructed based on the filtered *nPMI* scores. After removing rare words and filtering unimportant word pairs, there were 20,710 words and 134,072 edges in the Simplified Chinese semantic network and 12,137 words and 119,461 edges in the Traditional Chinese semantic network.

**Appendix D:**
**Determine the Threshold for Selecting Central Concepts**

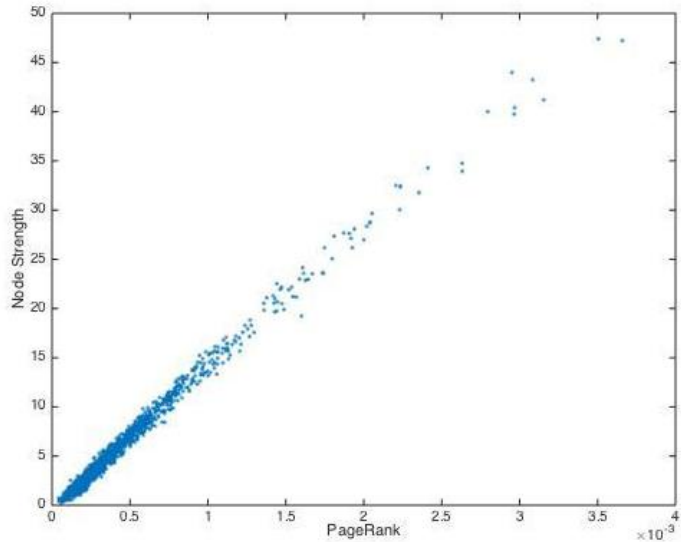

Figure D1 PageRank and Node Strength Scores of Nodes in the
Largest Concept Community of Simplified Chinese *Guanxi*
Semantic Network (N=3528)

When analyzing weighted network, we used node strength as edge weight. Node strength is the sum of all edges' weights of each node. Another measurement of centrality is PageRank, which not only measures at the number of links, but also considers the importance of the nodes it adheres to. Figure D1 shows the plot of the node strength and PageRank scores of all nodes in the largest concept community of Simplified Chinese *guanxi*, indicating a strong linear relation between the two measurements. We used node strength instead of PageRank because scores of PageRank tend to be relatively small (maximum value equals to 0.004), which adds difficulties to the comparison between nodes' centrality.

The distribution of node strength in a concept community follows the power low, meaning there is no natural cut-off value that could be used as the threshold of determining central concepts. For the best visualization purposes and considering the workload of translating central concepts in sub-communities, we set the threshold of selecting central concept as the $5^{th}$ percentile value of node strength, ranking in descending order.

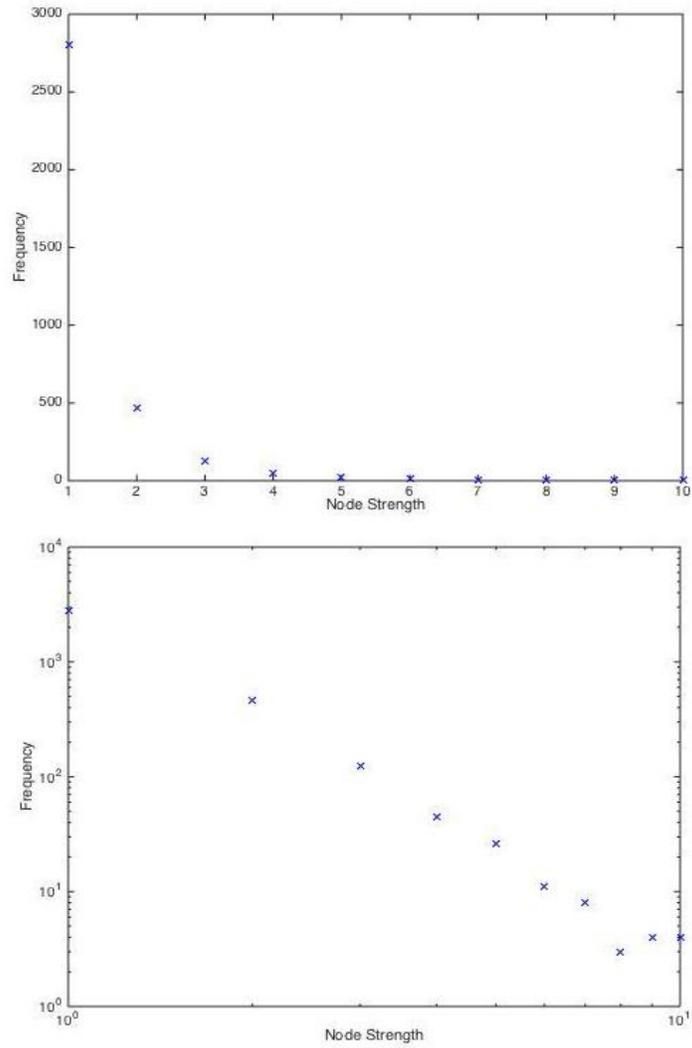

Figure D2 Distribution of Node Strength Scores in the Largest Concept Community of Simplified Chinese. (a) Distribution in linear scale. (b) Distribution in the logarithmic plot. Distribution of node strengths follows the power law.

**Appendix E:**
Semantic Clusters

Figure E1 Simplified Chinese Concept Community Concerning Intimate *guanxi* and Technology, the third Largest Concept Community within the Simplified Chinese *Guanxi* Semantic Network (N=815, 3.94% of all nodes). Only central concepts were translated into English.

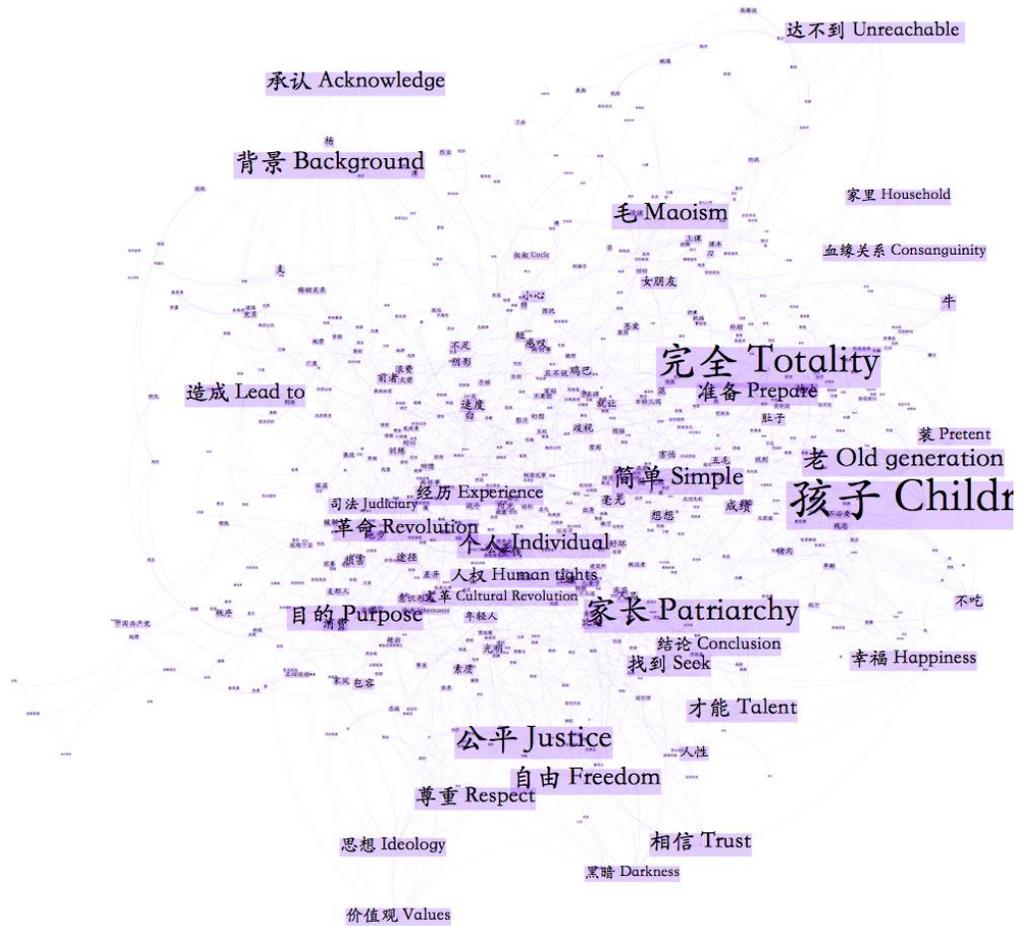

Figure E2 Simplified Chinese Concept Community Concerning *Guanxi* Traditions, fourth Largest Concept Community within the Simplified Chinese *Guanxi* Semantic Network (N=667, 3.21% of all nodes). Only central concepts were translated into English.

Figure E3 Traditional Chinese Concept Community Concerning Interpersonal Relationships. Largest Concept Community within the Traditional Chinese *Guanxi* Semantic Network (N=3224, 26.61% of all nodes). Only central concepts were translated into English

Figure E4 Simplified Chinese Concept Community Concerning Business and Government. Second Largest Concept Community within the Simplified Chinese *Guanxi* Semantic Network (N=987, 4.75% of all nodes). Only central concepts were translated into English

Figure E5 Traditional Chinese Concept Community Concerning Business. Second Largest Concept Community within the Simplified Chinese *Guanxi* Semantic Network (N=1444, 11.89% of all nodes). Only central concepts were translated into English.

Figure E6 Traditional Chinese Concept Community Concerning Politics. Fourth Largest Concept Community within the Simplified Chinese Guanxi Semantic Network (N=1313, 10.81% of all nodes). Only central concepts were translated into English.

Figure E7 Simplified Chinese Concept Community Concerning State Relations. Largest Concept Community within Simplified Chinese *Guanxi* Semantic Network (N=3528, 16.97% of all nodes). Only central concepts were translated into English.

Figure E8 Traditional Chinese Concept Community Concerning State Relations. Third Largest Concept Community within the Simplified Chinese *Guanxi* Semantic Network (N=1358, 11.18% of all nodes). Only central concepts were translated into English.